\documentclass[12pt, a4paper, twoside]{article}
\usepackage[a4paper, left=3cm,right=2cm]{geometry}
\usepackage[
      colorlinks=true,
      linkcolor=blue,
      urlcolor=blue,
      filecolor=blue,
      citecolor=red,
      pdfstartview=FitV,
      pdftitle={},
      pdfauthor={},
      pdfsubject={},
      pdfkeywords={},
      pdfpagemode=None,
      bookmarksopen=true
]{hyperref}
\usepackage{amsfonts}
\usepackage{amsmath}
\usepackage{setspace}
\usepackage{color}

\usepackage[utf8,applemac]{inputenc}
\usepackage{tensor}
\usepackage{cite}
\usepackage{tikz}
\usepackage{graphicx}
\usepackage[font=small, font+=it, format=hang, margin={1cm, 1cm}]{caption}
\graphicspath{{figure/}}

\usepackage{dcolumn}
\usepackage{bm}

\usepackage{subfig} 
\usepackage{enumitem} 

\usepackage[subfigure]{tocloft}

\newcommand{\bea}{\begin{eqnarray}}
\newcommand{\eea}{\end{eqnarray}}
\newcommand{\be}{\begin{equation}}
\newcommand{\ee}{\end{equation}}
\def\nn{\nonumber}

\def\p{\partial}
\def\eps{\epsilon}

\newcommand{\cI}{\mathcal{I}}

  \newcommand{\beqs}{\begin{eqnarray}}
\newcommand{\eeqs}{\end{eqnarray}}

\definecolor{GC}{rgb}{0,0.0,0.65}

\def\cI{\mathcal{I}}

\begin{document}


\setcounter{tocdepth}{2}

\begin{titlepage}

\begin{flushright}\vspace{-3cm}
{\small
\today }\end{flushright}
\vspace{0.5cm}

\begin{center}
{{ \LARGE{\bf{Asymptotically flat spacetimes \\ with $\text{BMS}_3$ symmetry \\ }}}} \vspace{5mm}

\centerline{\large{\bf{Geoffrey Comp\`{e}re\footnote{e-mail: gcompere@ulb.ac.be}, Adrien Fiorucci\footnote{e-mail:
afiorucc@ulb.ac.be}}}}

\vspace{2mm}
\normalsize
\bigskip\medskip
\textit{Universit\'{e} Libre de Bruxelles and International Solvay Institutes\\
CP 231, B-1050 Brussels, Belgium
}

\vspace{25mm}

\begin{abstract}
\noindent
{We construct the phase space of $3$-dimensional asymptotically flat spacetimes that forms the bulk metric representation of the BMS group consisting of both supertranslations and superrotations. The asymptotic symmetry group is a unique copy of the BMS group at both null infinities and spatial infinity. 
The BMS phase space obeys a notion of holographic causality and can be parametrized by boundary null fields. This automatically leads to the antipodal identification of bulk fields between past and future null infinity in the absence of a global conical defect. 
}

\end{abstract}


\end{center}

\end{titlepage}

\newpage
\begin{spacing}{1.125}
\tableofcontents
\end{spacing}
\vspace{0pt} 

\section{Introduction}

In Minkowski spacetime without gravity, all quantities of interest form representations of the Poincar\'e group \cite{1939AnMat..40..149W}. Instead, gravity with asymptotically flat boundary conditions in $4$ and $3$ dimensions admits the BMS group as asymptotic symmetry group \cite{Bondi:1962px,Sachs:1962wk}; \cite{Ashtekar:1996cd,Barnich:2006av}. The original BMS$_4$ group consists of the Poincar\'e group enhanced with an infinite-dimensional abelian subgroup of supertranslations. The BMS$_3$ group consists of both supertranslations and superrotations which enhance both the translation and Lorentz subgroups. The role of extended BMS symmetry in 4 dimensions that includes supertranslations as well superrotations is uncovering \cite{deBoer:2003vf,Barnich:2009se,Barnich:2010eb,Barnich:2011ct,Barnich:2011mi,Cachazo:2014fwa,Bern:2014oka,He:2014bga,Cachazo:2014dia,Geyer:2014lca,Kapec:2014opa,Broedel:2014fsa,Bern:2014vva,Campiglia:2014yka,Broedel:2014bza,Campiglia:2015yka,Flanagan:2015pxa,Barnich:2016lyg,Compere:2016jwb,Compere:2016hzt,Kapec:2016jld,Cheung:2016iub,Strominger:2016wns,Hawking:2016sgy,Conde:2016rom,He:2017fsb} but largely remains to be explored\footnote{Distinct definitions of $4d$ superrotations have been proposed in the literature. They are parametrized either by local conformal symmetries or by arbitrary diffeomorphisms on the two-sphere.}. The precise notion of boundary conditions and their multiple consequences for the infrared structure of gravity, is still a matter of ongoing research today, which is summarized in the lectures \cite{Strominger:2017zoo}.

In this work, we study the toy model of three-dimensional Einstein gravity as a starter for the physical four-dimensional case. The BMS$_3$ group has been originally defined at future null infinity $\mathcal I^+$ \cite{Ashtekar:1996cd,Barnich:2006av} (see also \cite{Barnich:2010eb,Bagchi:2012yk,Detournay:2014fva}) or, analogously, at past null infinity $\mathcal I^-$ where its representations have been largely understood \cite{Barnich:2012aw,Barnich:2012rz,Barnich:2014kra,Barnich:2015uva,Campoleoni:2016vsh,Oblak:2016eij,Carlip:2016lnw,Batlle:2017llu}. In a remarkable work, Strominger \cite{Strominger:2013jfa} argued that consistency of scattering in $4d$ asymptotically flat spacetimes requires a single copy of BMS acting both on ingoing and outgoing states, 
\bea
\text{BMS}_{\mathcal I^+} \times \text{BMS}_{\mathcal I^-} \rightarrow \text{BMS}.\label{id}
\eea
He showed that consistency with Lorentz covariance requires the generators of BMS to be antipodally identified between $\mathcal I^+_-$ the past of $\mathcal I^+$ and $\mathcal I^-_+$ the future of $\mathcal I^-$. The $3d$ case was argued similarly \cite{Prohazka:2017equ}. Nevertheless, no proof was given in terms of canonical methods. The purpose of this paper is to make precise the boundary conditions and phase space which lead to the identification \eqref{id} in the case of 3 dimensions. 

For that purpose, the hyperbolic foliation of asymptotically flat spacetimes  \cite{1978JMP....19.1542A,1982CMaPh..87...65B,deHaro:2000wj} (outside a given lighcone) is totally appropriate. Indeed, the asymptotic boundary of spacetime is conformal to 2-dimensional de Sitter spacetime and the relation between $\mathcal I^+_-$ and $\mathcal I^-_+$ can be understood in terms of wave propagation on the boundary de Sitter spacetime. This hyperbolic foliation is also a building block of holography in asymptotically flat spacetimes \cite{deBoer:2003vf}.

In the late stages of this work we received \cite{Campiglia:2017mua,Troessaert:2017jcm} which share some features with our analysis but concern distinct settings, namely electrodynamics \cite{Campiglia:2017mua} and $4d$ gravity \cite{Troessaert:2017jcm}.

The rest of the paper is organized as follows. We first construct the boundary conditions step by step in Section \ref{sec:BC}  and introduce the notion of holographic causality and boundary null field. We describe the phase space and asymptotic symmetry group resulting from a particular set of boundary conditions in Section \ref{sec:PS}. We also comment on the extension of symmetries in the bulk of spacetime following \cite{Barnich:2010eb,Compere:2014cna}. We summarize and further comment on the works \cite{Compere:2016jwb,Compere:2016hzt,Troessaert:2017jcm} in the final Section \ref{sec:CCL}. The Appendix \ref{app:Mink} contains a description of Minkowski spacetime and its Killing vectors in the hyperbolic foliation.

\section{Construction of boundary conditions}
\label{sec:BC}

\subsection{Solution space}
We first summarize the solution space of $3d$ Einstein gravity written in the hyperbolic foliation \cite{deHaro:2000wj,deBoer:2003vf}. 
The metric can be written exactly as
\bea
\label{hyp}
ds^2 = d\rho^2 + \left(\rho^2 h_{ab}^{(0)} +\rho h_{ab}^{(1)} + h_{ab}^{(2)}  \right)dx^a dx^b.
\eea
The bulk metric can be reconstructed from two holographic ingredients: the boundary metric $h^{(0)}_{ab}$ and the boundary stress-tensor $T^{ab}$  ($a,b$ take two values) as
\bea
h^{(1)}_{ab} &=& T_{ab} - h_{ab}^{(0)} h^{(0)}_{cd} T^{cd},\label{cons1}\\
h^{(2)}_{ab} &=& \frac{1}{4} h^{(1)}_{ac} h_{(0)}^{cd} h^{(1)}_{db}.\label{cons2}
\eea
Einstein's equations imply that the boundary metric is locally $dS_2$ with boundary Ricci scalar $R_{(0)} = 2$, and that the stress-tensor is conserved $\mathcal D_a T^{ab} = 0$. Here $\mathcal D_a$ is the boundary covariant derivative with respect to $h^{(0)}_{ab}$ and indices are  raised with the inverse boundary metric $h^{ab}_{(0)}$. The trace of the stress-tensor is not fixed, contrary to the analogous Fefferman-Graham expansion in $AdS_3$ \cite{Balasubramanian:1999re}.

\subsection{Diffeomorphism space}

Diffeomorphisms which preserve the solution space consist of boundary diffeomorphisms and Spi-supertranslations. The infinitesimal boundary diffeomorphisms are parametrized as $\xi ^\rho = 0$, $\xi^a = \xi_{(0)}^a(x^b)$ and infinitesimal Spi-supertranslations as
\bea
\xi^\rho &=& \omega(x^b),\nn \\
\xi^a &=&  \frac{1}{\rho} h_{(0)}^{ab}\mathcal D_b \omega -\frac{1}{2\rho^2} h_{(1)}^{ab} \mathcal D_b \omega + O(\rho^{-3}) .\label{Spitr}
\eea

The boundary metric does not transform under Spi-supertranslations. The stress-tensor transforms covariantly under boundary diffeomorphisms and it transforms under Spi-supertranslations as
\bea
\delta T_{ab} = 2  \mathcal D_a  \mathcal D_b \omega - 2h_{ab}^{(0)}  (\mathcal D^2  +1)\omega . \label{deltaT}
\eea
The Spi-supertranslations lead to the following transformation of the trace $T_a^a \equiv h_{(0)}^{ab}T_{ab}$, 
\bea
T_a^a \rightarrow T_a^a - 2(\mathcal D^2+2) \omega. 
\eea

\subsection{Holographic causality}

Gauge diffeomorphisms are not required to obey the principle of causality. Any gauge diffeomorphism can be performed at late times independently of the initial data. However, non-trivial diffeomorphisms which form the asymptotic symmetry group correspond to canonical transformations and therefore depend upon the initial data. In other words, non-trivial diffeomorphisms which are dictated by the asymptotic structure of spacetime obey a principle of boundary causality with respect to that asymptotic structure. 

In the hyperbolic foliation, bulk fields are decomposed in the asymptotic expansion with reference to the boundary $dS_2$ spacetime which provides with a boundary causality structure. It is expected that all bulk fields lead to boundary fields which propagate at or within the lightcone of $dS_2$. We will call such fields null and timelike boundary fields, respectively.  Boundary causality in particular requires that all boundary diffeomorphisms depend on at most two arbitrary functions on the circle which represent the initial data.

\subsection{Null boundary fields and antipodal map}

Null boundary fields, which propagate along the boundary lightcone, play a particular role which we now describe. Let us first write $dS_2$ in global coordinates as  
\bea
ds_{(0)}^2 =-d\tau^2+ \cosh^2 \tau d\phi^2 = \frac{-dT^2 + d\phi^2}{\cos^2 T} = - \frac{2 dx^+ dx^-}{\cos(x^+ + x^-) + 1}
\eea
where $\phi \sim \phi + 2\pi$. We define global conformal time $T$ as
\bea
\cosh\tau = \frac{1}{\cos T}, \qquad -\frac{\pi}{2} < T < \frac{\pi}{2}.\label{ch}
\eea
We set $T<0$ when $\tau < 0 $ and conversely $T > 0$ when $\tau > 0 $. The change of coordinates \eqref{ch} is then a bijection. The light-cone conformal coordinates are $x^\pm = T \pm \phi$. Both left and right moving light rays starting from the infinite past $T=-\frac{\pi}{2}$ reach the infinite future $T=+\frac{\pi}{2}$ at the \emph{antipodal point} on the circle. 

We identify $\mathcal I^+_-$, the past of future null infinity, with the future time infinity of the boundary hyperboloid, $\rho \rightarrow \infty$ and $T=\frac{\pi}{2}$. Similarly $\mathcal I^-_+$, the future of past null infinity, is identified with the past time infinity of the boundary hyperboloid, $\rho \rightarrow \infty$ and $T=-\frac{\pi}{2}$. Null boundary fields are therefore antipodally identified between $\mathcal I^+_-$ and $\mathcal I^-_+$. Timelike boundary fields on the contrary do not lead to such antipodal map. The boundary conditions for asymptotically flat spacetimes advocated in \cite{Strominger:2013jfa,Prohazka:2017equ} therefore requires null boundary fields. 

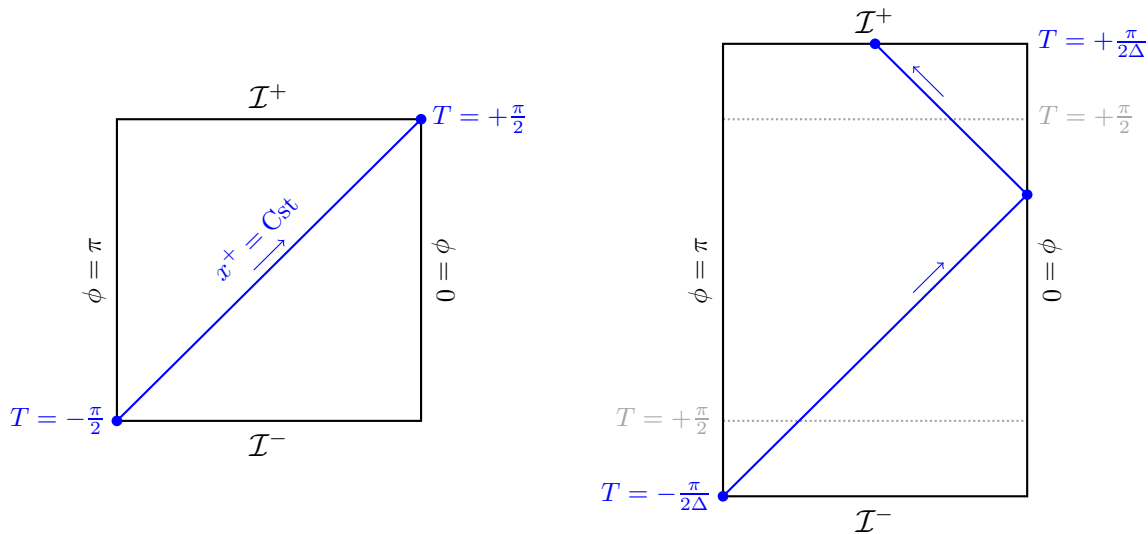
\begin{figure}[t!]
  \centering
  \subfloat[\textit{Without conical defect} ($\Delta = 1$).] {
	\begin{tikzpicture}[scale=1.0]
			\draw[white] (-3.5,-4.0) -- (-3.5,3.5) -- (3.5,3.5) -- (3.5,-4.0) -- cycle;
			\draw[thick] (-2,-2) -- (-2,2) -- (2,2) -- (2,-2) -- cycle;
			\node[below] at (0,-2) {$\cI^-$};
	       	\node[above] at (0, 2) {$\cI^+$};
	       	\node[above,rotate= 90] at (-2,0) {\footnotesize $\phi = \pi$};
	       	\node[above,rotate=270] at (2,0) {\footnotesize $\phi = 0$};
	       	\draw[thick,blue] (-2,-2) node[left] {{\footnotesize $T = -\frac{\pi}{2}$}} 
	       	-- (2,2) node[right] {{\footnotesize $T = +\frac{\pi}{2}$}};
	       	\fill[blue] (-2,-2) circle [radius=2pt];
	       	\fill[blue] ( 2, 2) circle [radius=2pt];	
	       	\draw[blue,->] (-0.2,0) -- (0.2,0.4);
	       	\node[blue,rotate= 45] at (-0.2,0.45) {\footnotesize $x^+ = \text{Cst}$};
	\end{tikzpicture}
} 
\quad
  \subfloat[\textit{With conical defect} ($0< \Delta < 1$).]{
	\begin{tikzpicture}[scale=1.0]
			\draw[white] (-3.5,-4.0) -- (-3.5,3.5) -- (3.5,3.5) -- (3.5,-4.0) -- cycle;
	       	\draw[thick, densely dotted,gray!70] (-2,-2) node[left]{{\footnotesize $T = +\frac{\pi}{2}$}} -- ( 2,-2);
	       	\draw[thick, densely dotted,gray!70] (-2, 2) -- ( 2, 2) node[right]{{\footnotesize $T = +\frac{\pi}{2}$}};
			\def\deltashift{1};
			\draw[thick] (-2,-2-\deltashift) -- (-2,2+\deltashift) -- (2,2+\deltashift) -- (2,-2-\deltashift) -- cycle;
			\node[below] at (0,-2-\deltashift) {$\cI^-$};
	       	\node[above] at (0, 2+\deltashift) {$\cI^+$};
	       	\node[above,rotate= 90] at (-2,0) {\footnotesize $\phi = \pi$};
	       	\node[above,rotate=270] at (2,0) {\footnotesize $\phi = 0$};
	       	\draw[thick,blue] (-2,-2-\deltashift) node[left] {{\footnotesize $T = -\frac{\pi}{2\Delta}$}} 
	       	-- (2,2-\deltashift) -- (0,2+\deltashift);
	       	\node[right,blue] at (2,3) {{\footnotesize $T = +\frac{\pi}{2\Delta}$}};
	       	\fill[blue] (-2,-2-\deltashift) circle [radius=2pt];
	       	\fill[blue] ( 0, 2+\deltashift) circle [radius=2pt];	
	       	\fill[blue] ( 2, 1) circle [radius=2pt];	
	       	\draw[blue,->] (0.5,-0.3) -- (0.9,0.1);
	       	\draw[blue,->] (0.9,2.3) -- (0.5,2.7);
	\end{tikzpicture}
}
  \caption{Penrose diagram illustrating null geodesic motion on the $dS_2$ boundary in the absence or presence of a global conical defect.}
  \label{fig:dS2Geodesics}
\end{figure}

Now, the possible presence of a global conical defect in $3$ dimensions \cite{Deser:1983tn} requires a special treatment\footnote{We will ignore in this work conical excesses which have an unbounded mass spectrum $M < -\frac{1}{8G} $ and cosmological solutions $M>0$ which have closed timelike curves.}. We keep the convention that $\phi \sim \phi + 2\pi$. The boundary metric of a conical defect of mass $-\frac{1}{8G} < M<0$ is then given by 
\bea
ds_{(0)}^2 = -d\tau^2+ \cosh^2 \tau \Delta^2 d\phi^2= \frac{\Delta^2 (-d T^2 + d\phi^2)}{\cos^2 (\Delta T)}  = - \frac{2 \Delta^2 dx^+ dx^-}{\cos(\Delta (x^+ + x^-) ) + 1}\label{conical}
\eea
where $x^\pm = T \pm \phi$ and 
\bea
 -\frac{\pi}{2\Delta} < T < \frac{\pi}{2\Delta}. 
\eea
The angular defect $\Delta=\sqrt{-8GM}$ obeys $0 < \Delta \leq 1$. Global conformal time $T$ now spans the interval $\frac{\pi}{\Delta}$ and null boundary fields are therefore identified between $\mathcal I^+_-$ and $\mathcal I^-_+$ up to the shift by $\frac{\pi}{\Delta}$ in their argument. This is illustrated in Figure \ref{fig:dS2Geodesics}.

\subsection{Existence of a variational principle}

The presymplectic potential form $\boldsymbol \Theta$ is defined from the variation of the action\footnote{More precisely, $\delta \boldsymbol L= \frac{\delta L}{\delta g_{\mu\nu}}\delta g_{\mu\nu}d^3 x+d\boldsymbol \Theta$ where $\boldsymbol \Theta = \frac{1}{2} \Theta^\mu \eps_{\mu\alpha\beta}dx^\alpha \wedge dx^\beta$ and $(16 \pi G \Theta_\mu) = \nabla^\nu \delta g_{\mu \nu} -\partial_\mu (g^{\alpha\beta}\delta g_{\alpha \beta})$.}.
The variational principle is well-defined if it exists a boundary counterterm $\mathbf B$ such that $\int_{\mathcal H} (\boldsymbol \Theta + \delta \mathbf B) =0$ at the asymptotic spatial boundary $\mathcal H$. A necessary and, assuming no topological obstruction, sufficient condition for the existence of the variational principle is the vanishing of the presymplectic structure $\boldsymbol \omega \equiv \delta \mathbf \Theta$ at the asymptotic boundary. 

Motivated by the role of null boundary fields, we impose that the boundary metric is written in boundary lightcone coordinates $(x^+,x^-)\sim (x^++2\pi,x^--2\pi)$, which obey
\bea
h_{++}^{(0)} = h_{--}^{(0)} = 0.\label{hpp}
\eea

The Lee-Wald presymplectic structure \cite{Lee:1990nz} pulled back on a constant hyperbolic slice can be then computed to be \bea
\omega^\rho[\delta g, \delta g ; g] = \frac{1}{32 \pi G} d^2 x\, \delta \sqrt{-h_{(0)}}  \wedge \delta T_a^a = \frac{1}{64 \pi G} d^2 x\, \delta \left( \sqrt{-h_{(0)}} R_{(0)} \right) \wedge \delta T_a^a .
\eea
(The second equality is trivial since $R_{(0)}=2$ but it suggests a relationship with the conformal anomaly structure of $3d$ gravity \cite{Henningson:1998gx}.) 

We identify three sets of boundary conditions together with \eqref{hpp}: 

\begin{itemize}[label=$\rhd$]
\item \textit{Dirichlet} conditions: 
\bea
\sqrt{-h_{(0)}} = \text{constant}.\label{DIR}
\eea
\item \textit{Neumann} conditions: 
\bea
T_a^a = \text{constant}.\label{NEU}
\eea  
\item Mixed conditions:
\bea
T_a^a = \mathcal F \left(\sqrt{-h_{(0)}}\right)
\eea
where $\mathcal F$ is a fixed function. 
\end{itemize}

Dirichlet boundary conditions together with \eqref{hpp} completely fix the boundary metric and therefore discard superrotations. We shall not consider such boundary conditions. Instead, we will consider Neumann boundary conditions. We will fully develop the phase space and asymptotic symmetry group in the next Section \ref{sec:PS}.

\section{The BMS phase space}
\label{sec:PS}

We now impose the Neumann boundary conditions \eqref{NEU} as
\bea
T^a_a = 0.\label{NEU2}
\eea
We impose that the boundary metric is described by boundary lightcone coordinates $(x^+,x^-)\sim (x^++2\pi,x^--2\pi)$ constrained as
\bea
h_{++}^{(0)} = h_{--}^{(0)} = 0.\label{hpp2}
\eea
Furthermore, we require asymptotic flatness at future null infinity in the sense of \cite{Barnich:2006av}. 

The boundary conditions ensure the existence of a variational principle. This promotes the solution space to a phase space which we describe next. 

\subsection{Phase space}

We consider the $3d$ metric in hyperbolic gauge \eqref{hyp}-\eqref{cons1}-\eqref{cons2}. Boundary diffeomorphisms which preserve  \eqref{hpp2} are  chiral null transformations 
\bea
x^+ \rightarrow X^+(x^+), \qquad x^- \rightarrow X^-(x^-).\label{idX}
\eea
Taking into account a global conical defect $\Delta$ as in \eqref{conical}, the general boundary metric can therefore be written as 
\bea
ds_{(0)}^2 = - \frac{2 \Delta^2 \p_+ X^+  \p_- X^- }{\cos (\Delta (X^+ + X^-)) + 1}dx^+ dx^-\label{ds0}
\eea
where $X^+ = X^+(x^+)$ and $X^-=X^-(x^-)$ are chiral functions and $0< \Delta \leq 1$. One has $\Delta=1$ in the absence of conical defect. Regularity of the boundary metric requires a non-vanishing determinant, which is equivalent to requiring the existence of real chiral null boundary fields $\Psi_+(x^+)$, $\Psi_-(x^-)$,
\bea
e^{\Psi_+} = \p_+ X^+,\qquad e^{\Psi_-} = \p_- X^-  .
\eea
After imposing the boundary condition \eqref{NEU2}, the conserved stress-tensor takes the generic form 
\bea
T_{++} &=& \Xi_+(x^+),\\
T_{+-} &=& 0,\nn \\
T_{--} &=&  \Xi_-(x^-) \nn 
\eea
where $\Xi_+ = \Xi_+(x^+)$ and $\Xi_-=\Xi_-(x^-)$ are null boundary fields. 

At this stage, the phase space is parameterized by four arbitrary functions $X^+(x^+)$, $X^-(x^-)$, $\Xi_+(x^+)$ and $\Xi_-(x^-)$, which are all chiral null boundary fields. The principle of boundary causality is therefore obeyed in this phase space. Since all boundary fields are null, all fields at future null infinity will be antipodally related to the ones at past null infinity in the absence of a global conical defect.

So far, the phase space metric is written in hyperbolic gauge as 
\bea
ds^2=d\rho^2 + 2h^{(0)}_{+-}\left( \rho \: dx^+ + \frac{\Xi_-(x^-)}{2h^{(0)}_{+-}} dx^-\right)\left( \rho \: dx^- + \frac{\Xi_+(x^+)}{2h^{(0)}_{+-}} dx^+\right) 
\eea
which is valid in the region $\rho \geq \rho_H(x^+,x^-)  $ defined by the locus where the determinant vanishes, 
\bea
\rho_H(x^+,x^-) = \frac{\sqrt{\Xi_+(x^+)\Xi_-(x^-)}}{2h^{(0)}_{+-}}.  
\eea
For future reference, we will refer to this locus at the BMS horizon in hyperbolic gauge.

Imposing asymptotic flatness at future null infinity requires to perform a gauge transformation to Bondi gauge. In our analysis, we found that no such gauge transformation exists to asymptotically flat spacetimes at future null infinity in the sense of \cite{Barnich:2006av}. In Minkowski spacetime, the denominator in \eqref{ds0} admits a zero at $T=\pm\frac{\pi}{2}$. We find that preserving this zero is essential for preserving asymptotic flatness at both null infinities. In the presence of a conical defect, the zero occurs at $\pm \frac{\pi}{2\Delta}$ and we need 
\bea
\cos \left[ \Delta \left( X^+ \left(\frac{\pi}{2\Delta} + \phi \right) + X^- \left(\frac{\pi}{2\Delta} - \phi\right) \right)\right] = -1.
\eea
We therefore require the following functional identification
\bea
X^-(x) = \frac{\pi}{\Delta} - X^+ \left(\frac{\pi}{\Delta} -x \right)  +\frac{2\pi}{\Delta}k , \qquad \forall x, \label{rel}
\eea
where $k  \in\mathbb Z$ labels disjoint BMS orbits. This identification is consistent with Lorentz transformations \eqref{Lorentz}. The change of coordinates to Bondi  gauge $(u,r,\varphi)$ at future null infinity takes the form
\bea
\rho &=& \sqrt{-2ur}\left( 1+O(r^{-1}) \right), \\
T &=& \frac{\pi}{2\Delta}-\sqrt{2}\frac{\sqrt{-u}}{\sqrt{r}}\left( 1+O(r^{-1}) \right),\label{chg}\\
\phi &=& \varphi +O(r^{-2}),
\eea
where subleading terms are uniquely fixed. Up to the rescaling of $T$ necessary to take the global conical defect into account, the leading order transformation of \eqref{chg} matches the one of Minkowski spacetime \eqref{chM}, which ensures that the coordinate transformation is a gauge transformation. The metric reads as
\bea
ds^2 = \Theta(\varphi) du^2 -2 du dr + 2 \left[ \Xi(\varphi) + \frac{u}{2} \p_\phi \Theta (\varphi) \right]dud\varphi + r^2 d\varphi^2
\eea
where 
\begin{align}
\Xi(\varphi) &= \frac{1}{2} \left[ \Xi_+ \left(\frac{\pi}{2\Delta}+\varphi\right) -\Xi_-\left(\frac{\pi}{2\Delta} - \varphi\right) \right], \nn \\
\Theta(\varphi) &= \left(\p_\varphi \Psi(\varphi) \right)^2 - 2 \p_\varphi^2 \Psi(\varphi) -\Delta^2 e^{2\Psi(\varphi)},\qquad e^{\Psi(\varphi)} \equiv \p_\varphi X^+ \left(\frac{\pi}{2\Delta}+\varphi \right).  \label{idd}
\end{align}
This is exactly the metric (3.14) of \cite{Barnich:2010eb} with their $\varphi = 0$ or, equivalently, the metric (47) of \cite{Barnich:2012aw}. The expression for $\Theta(\varphi)$ in \eqref{idd} is recognized as the Liouville stress-tensor of the field $\Psi(\varphi)$. Even though the original hyperbolic gauge is valid only for $u<0$, Bondi gauge also covers $u \geq 0$. \\
Similarly, we also obtain that the phase space is  asymptotically flat at past null infinity in the sense of \cite{Barnich:2006av} after imposing the periodicity requirements
\begin{alignat}{3}
X^+(x) &= X^+ \left(x + \frac{2\pi}{\Delta} \right)+2\pi \hat k, &&\quad \forall x, \\
\Xi_-(x)&= \Xi_- \left(x+\frac{2\pi }{\Delta} \right), &&\quad \forall x,
\end{alignat}
where $\hat k \in \mathbb Z$ again labels distinct orbits. Bondi gauge $(v,r,\varphi)$ at past null infinity is reached upon imposing (with additional details given)
\bea
\rho &=& \sqrt{2vr}\left[1+\frac{v}{4r}\Theta\left(\varphi-\frac{\pi}{\Delta}\right)+O(r^{-2})\right], \\
T &=& -\frac{\pi}{2\Delta}+\sqrt{2}\frac{\sqrt{v}}{\sqrt{r}}\left[1-\frac{v}{12r} \Theta\left(\varphi-\frac{\pi}{\Delta}\right)+O(r^{-2}) \right],\\
\phi &=& \varphi -\frac{v}{r^2}\left[\Xi\left(\varphi -\frac{\pi}{\Delta}\right) -\frac{v}{12}\partial_\varphi\Theta\left(\varphi-\frac{\pi}{\Delta}\right)\right] +O(r^{-3}).
\eea
The metric then reads as
\bea
ds^2 = \Theta\left(\varphi-\frac{\pi}{\Delta}\right) dv^2 +2 dv dr - 2 \left[ \Xi\left(\varphi-\frac{\pi}{\Delta}\right) - \frac{v}{2} \p_\phi \Theta \left(\varphi-\frac{\pi}{\Delta}\right) \right]dvd\varphi + r^2 d\varphi^2
\eea
where the phase space canonical variables are again defined exactly as \eqref{idd}.  
Bondi gauge is valid for all $v$.

In summary, the phase space is parametrized by the three independent null boundary fields $X^+(x^+)$, $\Xi_+(x^+)$ and $\Xi_-(x^-)$. However, only one combination of $\Xi_+$ and $\Xi_-$ appears at future and past null infinity. This suggests that other supertranslations are pure gauge, which we will confirm in the following. The supertranslation and superrotation fields at $\mathcal I^-$ are related to the fields defined at $\mathcal I^+$ by a shift of $\pi$ over the conical deficit angle $\Delta$. In the absence of conical defect, the fields are antipodally identified.


\subsection{Asymptotic symmetries and charges}

\noindent \underline{\textit{Superrotations}:}\vspace{10pt}

The boundary diffeomorphisms which preserve the phase space are the superrotations 
\bea
\xi_R &=& R^+ (x^+) \p_+ +R^-(x^-)\p_-,\qquad R^-(x^-) \equiv - R^+ \left(\frac{\pi}{\Delta} - x^-\right)
\eea
labelled by an arbitrary function $R^+(\cdot )$ on $S^1$. The left and right moving chiral components of superrotations are related as a consequence of the boundary conditions, see \eqref{rel} above. In a Fourier decomposition, the 3 lowest modes correspond to Lorentz transformations \eqref{Lorentz}. In Bondi gauge, a superrotation takes the standard form 
\bea
\xi_R = ( R(\varphi) + O(r^{-1})) \p_\varphi + (u R'(\varphi) + O(r^{-1})) \p_u + ( -r R'(\varphi) +O(r^0) ) \p_r\label{xiRB}
\eea
where we identified the generator at future null infinity as 
\bea
R(\varphi) = R^+\left(\frac{\pi}{2\Delta}+\varphi\right).\label{defRR}
\eea
Superrotations preserve the phase space and only transform the canonical fields. Their transformation follows from the action of the diffeomorphism and is obtained as 
\bea
\delta_R X^+ &=& R^+ \p_+ X^+ ,\qquad \delta_R \Xi_+ = 2 \Xi_+ \p_+ R^+ + R^+ \p_+ \Xi_+,\\
\delta_R X^- &=& R^- \p_- X^-, \qquad \delta_R \Xi_- = 2 \Xi_- \p_- R^- + R^-\p_- \Xi_-.
\eea

Let us quickly summarize the definition of canonical charges which we need. The canonical charges associated with asymptotic symmetries are given by the Iyer-Wald charges \cite{Wald:1993nt}\footnote{Alternatively, the canonical charges  are given by the Barnich-Brandt charges \cite{Barnich:2001jy}. Both definitions agree because the charge difference, given in (E.14) of \cite{Barnich:2007bf}, identically vanishes in hyperbolic gauge.}. The infinitesimal surface charge one-form takes the form $\boldsymbol k_\xi[\delta g ; g] = \delta \boldsymbol Q_\xi[g] - \boldsymbol Q_{\delta \xi}[g] - \xi \cdot \boldsymbol \Theta [\delta g ; g]$ where $\boldsymbol Q_\xi[g]$ is the Komar form. Such charges are integrable (\textit{i.e.} are the variation of a well-defined charge) if and only if the integrability conditions $\int_S \delta_1 \boldsymbol k_\xi[\delta_2 ; g] - (1 \leftrightarrow 2) = 0$ are obeyed, see \cite{Barnich:2007bf} for a general discussion. We will use the orientation $\epsilon_{\rho + -}=-1$.

It turns out that the superrotation charges, also called \textit{super-Lorentz charges} $\mathcal J$, are integrable on the phase space. They are given by
\bea
\mathcal J \equiv \mathcal Q_R = \frac{1}{16\pi G}\int_0^{2\pi} d\phi \; \left( R^+(x^+) \Xi_+(x^+) +R^-(x^-) \Xi_-(x^-) \right).
\eea
The superrotations charges are clearly conserved. Upon evaluating them at $t=0$ and using \eqref{defRR}-\eqref{idd} we can rewrite them as
\bea
\mathcal Q_R =  \frac{1}{8\pi G}\int_0^{2\pi} d\varphi \;  R(\varphi) \Xi(\varphi) . \label{chR}
\eea

\noindent \underline{\textit{Supertranslations}:}\vspace{10pt}

The Spi-supertranslations \eqref{Spitr} which preserve the phase space are parametrized by the function $\omega$ which obeys $(\mathcal D^2 +2)\omega = 0$. This is a second-order partial differential equation on $dS_2$. After some algebra, we find remarkable that the explicit general solution can be obtained in closed form as 
\bea
\omega =\Delta (S_+ + S_-) \tan\left( \Delta \frac{X^+ + X^-}{2} \right)+ \frac{\p_+ S_+}{\p_+ X^+} + \frac{\p_- S_-}{\p_- X^-} \label{defo}
\eea
where $S^+(x^+)$ and $S^-(x^-)$ are arbitrary chiral functions. With hindsight, it turns out to be convenient to perform the following field redefinition 
\bea
S_+(x^+) = -\frac{1}{2}T^+(x^+) \p_+ X^+(x^+),\qquad S_-(x^-) = -\frac{1}{2}T^-(x^-) \p_- X^-(x^-). \label{defT}
\eea
A supertranslation vector $\xi_T = \xi_T(\omega)$ is therefore labelled by two chiral functions $T^+ =T^+(x^+)$ and $T^ - =T^-(x^-)$. 
In Bondi gauge at future null infinity, the supertranslations take the standard form 
\bea
\xi_T = T(\varphi)\p_u + O(r^{0}) \p_r + O(r^{-1})\p_\varphi\label{xiTB}
\eea
where 
\bea
T(\varphi) =\frac{1}{2} \left[  T^+ \left(\frac{\pi}{2\Delta}+\varphi\right) + T^- \left(\frac{\pi}{2\Delta}-\varphi\right) \right].\label{TT}
\eea
In Bondi gauge at past null infinity, we find the expected generator,
\bea
\xi_T = T\left(\varphi+\frac{\pi}{\Delta}\right)\p_v + O(r^{0}) \p_r + O(r^{-1})\p_\varphi .
\eea

Considering $S_\pm$ to be field-dependent in \eqref{defT} in terms of $T^\pm$ that are field independent corresponds to choosing an integrating factor for defining the charges \cite{Barnich:2007bf}. The supertranslation charges, also called \textit{supermomenta} $\mathcal P$, are integrable and given by 
\bea
\mathcal P\equiv \mathcal Q_T = \frac{1}{32\pi G}\int_0^{2\pi} d\phi \; \left[ T^+(x^+) (\Theta_+(x^+) +1)+T^-(x^-) (\Theta_-(x^-)+1 )\right]
\eea
where we defined
\bea
\Theta_+(x^+) & \equiv & (\p_+ \Psi^+)^2 - 2\p_+^2\Psi_+-\Delta^2 e^{2\Psi_+},\qquad e^{\Psi_+} \equiv \p_+ X^+(x^+), \\
\Theta_-(x^-) & \equiv & (\p_- \Psi^-)^2 - 2\p_-^2\Psi_--\Delta^2 e^{2\Psi_-},\qquad e^{\Psi_-} \equiv \p_- X^-(x^-). 
\eea
The charges are conserved. Here we conventionally fixed the normalization of the charges such that Minkowski has vanishing charges. After noting $\Theta_+(\phi+\frac{\pi}{2\Delta}) = \Theta_-(-\phi+\frac{\pi}{2\Delta})=\Theta(\phi)$ where $\Theta(\phi)$ is defined in \eqref{idd}, we can rewrite the supertranslation charges as
\bea
\mathcal Q_T = \frac{1}{16\pi G}\int_0^{2\pi} d\varphi \; T(\varphi) (\Theta(\varphi) +1).\label{chT}
\eea

Since only the combination \eqref{TT} is associated to non-trivial supertranslation charges, we can discard the other supertranslations which are pure gauge. We could gauge fix the gauge supertranslations by imposing the functional dependence
\bea
T^+(x) = T^- \left(\frac{\pi}{\Delta} -x\right) ,\qquad \forall x.\label{idT}
\eea
This constraint is obeyed for the 3 translations of Minkowski spacetime, see Appendix \ref{app:Mink}. The gauge fixed phase space is therefore still consistent with the Poincar\'e group.

\subsection{BMS algebra}

The variation of the stress-tensor under Spi-supertranslations \eqref{deltaT} allows to deduce the transformation properties of $\Xi_+$ and $\Xi_-$ under supertranslations. We find  
\bea
\delta_T \Xi_+ &=& \Theta_+ \p_+ T^+ +\frac{1}{2} T^+ \p_+ \Theta_+ -\p_+^3 T^+ ,\\
\delta_T \Xi_-&=& \Theta_- \p_- T^- +\frac{1}{2} T^- \p_- \Theta_- -\p_-^3 T^-. 
\eea

Using the identification \eqref{idd} we also deduce the transformation properties of the fields at future null infinity under supertranslations and superrotations
\bea
\delta_{T,R} \Xi &=& 2 \Xi \p_\varphi R + R \p_\varphi \Xi +  \Theta \p_\varphi T + \frac{1}{2} T \p_\varphi \Theta - \p_\varphi^3 T ,\nn \\
\delta_{T,R} \Theta &=& R \p_\varphi \Theta + 2\p_\varphi R \Theta - 2\p_\varphi^3 R, \label{eq:4}
\eea
which exactly reproduces (3.15) of \cite{Barnich:2010eb} once we align our conventions (we use $\delta_\xi g_{\mu\nu}= +\mathcal L_\xi g_{\mu\nu}$). 
Note that the transformations \eqref{eq:4} can be written more fundamentally as 
\bea
\delta_{T,R}C  &=& T(\varphi) + R(\varphi) C'(\varphi) - C(\varphi)R'(\varphi),\\
\delta_{T,R}\Psi &=& R(\varphi) \Psi'(\varphi)+R'(\varphi),
\eea
once we recognize\footnote{We thank R. Oliveri for helping to derive the second relation.}
\bea
\Theta[\Psi ] &=& ( \Psi'(\varphi))^2 - 2 \Psi'' (\varphi)+8 G M e^{2\Psi(\varphi)},\\
\Xi[C,\Psi] &=& \Theta(\varphi) C'(\varphi) - C'''(\varphi)+4 G J e^{2 \Psi(\varphi)} + \frac{1}{2} \Theta'(\varphi) C(\varphi)
\eea
in terms of the mass $M$, angular momentum $J$, supertranslation field $C(\varphi)$ and Liouville field $\Psi(\varphi)$.

The commutation relations of the asymptotic symmetries under the Lie bracket are easily obtained by computing the Lie bracket of the supertranslation and superrotation generators \eqref{xiTB} and \eqref{xiRB}. Alternatively, one could compose two transformations of the type \eqref{eq:4} and identify the resulting transformation. For simplicity, we concentrate on the non-trivial asymptotic symmetries by imposing the gauge fixing condition \eqref{idT}. We label by $R_n$ the superrotation vector $\xi_R$ where we select the Fourier mode $R(\phi)=e^{i n \phi}$ and we label by $T_n$ the supertranslation vector $\xi_T$ where we select the Fourier mode $T(\phi)=e^{i n \phi}$. We obtain the Lie bracket 
\bea
i [T_m, T_n] &=& 0 ,\\
i [R_m , T_n ]&=& (m-n) T_{m+n}, \\
i [R_m , R_n] &=& (m-n)R_{m+n}. 
\eea
This is precisely the BMS$_3$ algebra \cite{Ashtekar:1996cd,Barnich:2006av}. 

The charge bracket is defined as 
\bea
\{ Q_{\xi_1} , Q_{\xi_2} \} = \delta_{\xi_2} Q_{\xi_1} .\label{bra}
\eea
Since the charges are integrable, the charge bracket forms an algebra isomorphic to the symmetry algebra up to central extensions \cite{Barnich:2001jy,Barnich:2007bf}. We denote the modes of the supermomenta as $\mathcal P_m = \mathcal Q_{T_m}$ and the modes of the superrotation charges as $\mathcal J_m = \mathcal Q_{R_m}$. Using \eqref{chT}-\eqref{chR} and \eqref{eq:4} we find 
\bea
i \{ \mathcal P_m, \mathcal P_n \} &=& 0 ,\\
i \{ \mathcal J_m , \mathcal P_n \}&=& (m-n) \mathcal P_{m+n} + \frac{1}{4G}m(m^2-1)\delta_{m+n,0},\label{BMSA}\\
i \{ \mathcal J_m , \mathcal J_n \} &=& (m-n)\mathcal J_{m+n}. 
\eea
This is exactly the BMS$_3$ charge algebra \cite{Barnich:2006av}.

\subsection{Symplectic symmetries}

The BMS algebra is realized in the bulk of spacetime in two complementary senses.
 
First, it exists a modified Lie bracket \cite{Barnich:2010eb} such that the algebra of symmetries holds in the bulk of spacetime. This modified Lie bracket is defined as 
\bea
[\xi_1,\xi_2]_M =[\xi_1,\xi_2] - \delta^g_{\xi_1}\xi_2 +\delta_{\xi_2}^g \xi_1,
\eea
where $\delta^g_{\xi_1}\xi_2$ denotes the variation of $\xi_2$ caused by the Lie dragging along $\xi_1$ of the metric contained in the definition of $\xi_2$. Such modified terms are relevant for supertranslations since subleading terms depend upon the metric, see \eqref{Spitr}.  

Second, the symplectic structure on the phase space identically vanish locally (except at possible sources): 
\bea
\boldsymbol \omega [\delta g, \delta g ;g] =0. \label{omega}
\eea
All variations along the symmetries therefore lead to conserved charges in the bulk of spacetime as a consequence of the generalized Noether theorem \cite{Wald:1993nt,Barnich:2000zw,Barnich:2001jy}. This theorem states that the infinitesimal canonical charge $\boldsymbol k_\xi[\delta g ; g]$ is related to the symplectic form $\boldsymbol \omega$ on-shell and for linearized on-shell perturbations as 
\bea
\boldsymbol \omega[\delta g, \mathcal L_\xi g ;g] = d \boldsymbol k_\xi[\delta g ; g].
\eea
The canonical charges are therefore conserved in the bulk of spacetime as a consequence of \eqref{omega} and Stokes' theorem. Asymptotic symmetries  therefore extend to bulk symmetries, as observed in \cite{Compere:2014cna}. Such symmetries have been coined as symplectic symmetries \cite{Compere:2015knw}. They obey the BMS algebra \eqref{BMSA} under the standard bracket \eqref{bra} in the bulk of spacetime. This ends our analysis of the phase space.

\section{Summary and comments}
\label{sec:CCL}

We found the boundary conditions which ensure the existence of a single copy of the BMS group as symmetry group of asymptotically flat spacetimes. This symmetry group is asymptotic but also symplectic, \textit{i.e.} valid in the bulk of spacetime. A point in the phase space is a metric which carries definite supermomenta and super Lorentz charges. Since these charges are conserved at both null infinities, spatial infinity and in the bulk of spacetime, the entire spacetime carries the charges. As a consequence, there ought to be a bulk source for these charges which can be traced to the location where the finite solution-generating diffeomorphism which turns on the charges  has vanishing determinant (this does not apply for the zero modes, mass and angular momentum, which correspond to conical defects). This locus, coined as the BMS horizon \cite{Compere:2016jwb,Compere:2016hzt}, is gauge dependent but has to exist because of charge conservation.

In gravity, observables defined as surface integrals obey the screening property:  details of the sources are hidden beyond the surface which encloses them all. For example, the total angle deficit with respect to $2\pi$ in a large annulus around a disk gives the total angle deficit of all individual conical defects in the disk. The same property holds for BMS charges: sources within each BMS horizon will each contribute to the total charge of a surface enclosing all sources. This is a consequence of the linear superposition property of symplectic charges \cite{Compere:2015knw} in complete analogue with the linear superposition property of Killing charges \cite{Barnich:2003xg}. Notably, this implies that the relative charge of two individual sources is independent upon the asymptotic charges.

The main aim of this analysis was to obtain a complete description of the $\text{BMS}$ group in $3$ dimensions as a starter for $4$ dimensions. In $3d$, the phase space and symmetries are described analytically in closed form. Our boundary conditions lead to supertranslations and superrotations defined as boundary null fields which automatically imply their antipodal map property in the absence of a global conical defect. In the presence of a conical deficit angle $\Delta$, we obtained that the fields at past null infinity with respect to future null infinity are instead shifted by $\pi/\Delta$. The antipodal map property of $4d$ supertranslations was made analogously in \cite{Troessaert:2017jcm} after imposing consistent boundary conditions at spatial infinity \cite{Compere:2011ve} and upon imposing asymptotic flatness at null infinity. 

We concentrated in this work on specific boundary conditions but other consistent boundary conditions may be possible, some of which may contain timelike boundary fields. The definition of supertranslations at spatial infinity used in \cite{Compere:2016jwb,Compere:2016hzt} uses timelike boundary fields, which do not obey the antipodal map. A complete classification of boundary conditions at spatial infinity is certainly desirable in order to settle the uniqueness, or not, of the asymptotic structure of spacetimes without cosmological constant. Alternative boundary conditions have recently been considered at null infinity \cite{Detournay:2016sfv,Grumiller:2017sjh} but remain to be linked to spatial and past null infinity.

\section*{Acknowledgements}

We are grateful to Glenn Barnich, Jiang Long and Roberto Oliveri for useful conversations. G.C. is a Research Associate of the Fonds de la Recherche Scientifique F.R.S.-FNRS (Belgium) and he acknowledges the current support of the ERC Starting Grant 335146 ``HoloBHC".

\appendix

\section{Minkowski spacetime in the hyperbolic foliation}
\label{app:Mink}

Minkowski spacetime can be foliated by hyperboloids outside the lightcone centered at $\rho=0$ using 
\bea
t &= &\rho \sinh\tau, \qquad x= \rho\cosh\tau \cos\phi,\qquad y = \rho \cosh\tau \sin\phi .
\eea
The metric is
\bea
ds^2 = d\rho^2 + \rho^2 (- d\tau^2 + \cosh^2\tau d\phi^2) = d\rho^2 + \frac{\rho^2}{\cos^2 T}(-dT^2 + d\phi^2),
\eea
where $\phi \sim \phi+2\pi$, $-\infty < \tau < \infty$, $-\pi/2 < T< \pi/2$ and we defined $x^\pm = T\pm \phi$.

Rotations and boosts are given by
\begin{alignat}{4}
\xi_{\text{rot}} &= -y \p_x &&+ x \p_y &&= \p_\phi = \p_+ - \p_-,\nn\\
\xi_{\text{boost},x} &= \phantom{-}x \p_t &&+ t \p_x &&= \cos\phi \p_\tau -\tanh\tau \sin\phi \p_\phi = \cos x^+ \p_+ + \cos x^- \p_-, \label{Lorentz}\\
\xi_{\text{boost},y} &= \phantom{-}y \p_t &&+ t \p_y  &&= \sin\phi \p_\tau +\tanh\tau \cos\phi \p_\phi = \sin x^+ \p_+ - \sin x^- \p_-. \nn
\end{alignat}

Time and spatial translations are given by
\bea
\p_t &=& -\sinh \tau \p_\rho + \frac{\cosh\tau}{\rho}\p_\tau \\
&=& -\tan \left(\frac{x^++x^-}{2}\right)\p_\rho + \frac{1}{\rho}(\p_+ + \p_-),\\
\p_x &=& \cosh\tau \cos \phi \p_\rho -\frac{\sinh\tau\cos\phi}{\rho}\p_\tau - \frac{\sin\phi}{\rho \cosh\tau } \p_\phi \\
&=& \frac{\cos (\frac{x^+-x^-}{2})}{\cos (\frac{x^++x^-}{2})}\p_\rho -\frac{1}{\rho} (\sin x^+ \p_+ + \sin x^- \p_-), \\
\p_y &=& \cosh\tau \sin \phi\p_\rho  -\frac{\sinh\tau\sin\phi}{\rho}\p_\tau + \frac{\cos\phi}{\rho \cosh\tau } \p_\phi \\
&=&  \frac{\sin (\frac{x^+-x^-}{2})}{\cos (\frac{x^++x^-}{2})}\p_\rho +\frac{1}{\rho} (\cos x^+ \p_+ -\cos x^- \p_-).
\eea
We can therefore read $\omega$ in the definition \eqref{Spitr}. Translations correspond via \eqref{defo}-\eqref{defT} to $(T^+,T^-) =(1,1)$ for time translations, $(T ^+,T^-)=(-\sin x^+,-\sin x^-)$ for $x$ translations and $(T ^+,T^-)=(\cos x^+, -\cos x^-)$ for $y$ translations which are all consistent with \eqref{idT}.

The gauge transformation to Bondi gauge at future null infinity 
\bea
ds^2 = -du^2 -2 dudr +r^2 d\phi^2
\eea
 is defined for $u<0$ as
\bea
u &=& -\rho e^{-\tau} = \rho (\tan T - \cos^{-1} T),\\
r&=& \rho \cosh \tau = \frac{\rho}{\cos T},
\eea
and its inverse is
\bea
\rho &=& \sqrt{(-u)(2r+u)} = \sqrt{2}\sqrt{-u}\sqrt{r} + O(r^{-1/2}), \\
T &=& \arctan \left[ \frac{r+u}{\sqrt{(-u) (2r+u)}} \right] = \frac{\pi}{2}-\sqrt{2}\frac{\sqrt{-u}}{\sqrt{r}}+O(r^{-3/2}).\label{chM}
\eea
The gauge transformation to Bondi gauge at past null infinity
\bea
ds^2 = -dv^2 +2 dvdr +r^2 d\phi^2
\eea
is defined for $v > 0$ as
\bea
v &=& \rho e^{\tau} = \rho (\tan T + \cos^{-1} T),\\
r&=& \rho \cosh \tau = \frac{\rho}{\cos T},
\eea
and its inverse is
\bea
\rho &=& \sqrt{v(2r-v)} = \sqrt{2}\sqrt{v}\sqrt{r} + O(r^{-1/2}) ,\\
T &=& -\arctan \left[ \frac{r-v}{\sqrt{v (2r-v)}} \right] = -\frac{\pi}{2}+\sqrt{2}\frac{\sqrt{v}}{\sqrt{r}}+O(r^{-3/2}).
\eea

\newpage


\providecommand{\href}[2]{#2}\begingroup\raggedright\endgroup

\end{document}